\documentclass[aps]{revtex4}
\usepackage[colorlinks=true]{hyperref}  
\usepackage{graphicx,color}
\usepackage{latexsym}
\usepackage{amsmath}
\usepackage{amsfonts}
\usepackage{amssymb}
\usepackage[latin1]{inputenc}

\newcommand{\be}{\begin{equation}}
\newcommand{\ee}{\end{equation}}
\newcommand{\ben}{\begin{eqnarray}}
\newcommand{\een}{\end{eqnarray}}

\begin{document}
\title{Self-tuning Horava-Lifshitz's  Universe}
\author{D. Bazeia$^{a,b,c}$, F.A. Brito,$^{c}$ and F.G. Costa,$^{b,d}$}
\email{bazeia@fisica.ufpb.br, fabrito@df.ufcg.edu.br, geraldo.costa@ifrn.edu.br}
\affiliation{$^a$Instituto de F\'\i sica, Universidade de S\~ao Paulo, 05314-970, S\~ao Paulo, SP, Brazil\\
$^{b}$Departamento de  F\'\i sica, Universidade Federal da  Para\'\i ba, 58051-970 Jo\~ao Pessoa, Para\'\i ba, Brazil\\
$^{c}$Departamento de F\'\i sica, Universidade Federal de Campina Grande, Caixa Postal 10071, 58109-970 Campina Grande, Para\'\i ba, Brazil\\ 
$^{d}$Instituto Federal do Rio Grande do Norte-IFRN, 59380-000 Currais Novos, Rio Grande do Norte, Brazil}

\begin{abstract} 
In this paper we show that by considering a slightly modified Horava-Lifshitz gravity one can find self-tuning cosmological scenarios. 
Since in the proposed model the cosmological constant cancels out, then 
regardless of its  value it cannot produce any gravitational effect. As a consequence this removes the gravitational effect of the vacuum energy on the expansion of the Universe. Accelerating scenarios, however, can be found 
through the presence of conventional and non-conventional fluids.

\end{abstract}
\maketitle
\pretolerance10000



\section{Introduction}

One of the most longstanding problem in physics is the cosmological constant problem \cite{weinberg}. Since it is related to the vacuum energy it is expected to be computed from particle physics, as in the realm of the standard model. However the theoretically computed vacuum energy is  much bigger than its natural value. In the most attempts to solve the problem one mainly considers improving of the calculation by putting some new symmetries to make the vacuum energy much smaller than its natural value or even forcing cancelation such as in supersymmetric  extensions. 

However, one can also keep the large vacuum energy, but changing the gravitational dynamics such that the gravitational effect of the cosmological constant is very small or even zero. This has been already discussed in {\it self-tuning scenarios}. The former case was addressed  in self-tuning branes \cite{ArkaniHamed:2000eg, Kachru:2000xs}, whereas the latter was considered in braneworlds \cite{carr}, where the cosmological constant is exactly canceled out due to `natural' modification of the induced four-dimensional Friedmann equation.

One of the main challenges in constructing such self-tuning scenarios is the difficulty of changing the gravitational dynamics in such a way the effects of the cosmological constant can be removed, but the conventional cosmology or at least part of it should be recovered in the modified Friedmann equation. For recent works addressing the cosmological constant problem by changing the gravitational dynamics see \cite{padilla,Garattini:2009ns}.

In this paper, we apply the recently proposed Horava-Lifshitz (HL) gravity \cite{hor} to address the self-tuning issue. This theory of gravity scales the space and time in an anisotropic way and present UV-completion in the Lifshitz point $z=3$. The IR regime is also modified by a cosmological constant and a potential fixed by the detailed balance. We have shown that HL gravity when properly modified can precisely cancel the cosmological constant.   The cosmological constant does not contribute into modified Friedmann equation, but the conventional cosmology can be easily recovered. The aforementioned modification has a close relationship with the scalar graviton free Horava-Melby-Thompson (HMT) theory \cite{Horava:2010zj} --- See also \cite{daSilva:2010bm}.  As we shall see later, our proposed deformed HL gravity can be seen as the HMT gravity in the limit of null gauge field and the Newtonian prepotential developing a properly chosen {\it non zero} value.

Other cosmological issues have already been considered in the recent literature both in HL gravity \cite{Brandenberger:2009yt}
and HMT gravity \cite{Wang:2010wi}.

The paper is organized as follows. In Sec.~\ref{s1} we present our set up to produce self-tuning in the Horava-Lifshitz gravity. In Sec.~\ref{HMT-sec} we show how to relate our deformed HL gravity to HMT gravity. In Sec.~\ref{s2} and Sec.~\ref{s3} we show examples where the cosmological constant can be canceled out while many scenarios of conventional cosmology is recovered. Finally, in Sec.~\ref{conc} we make all final comments.

\section{Construction out of Horava-Lifshitz Gravity}
\label{s1}

Based on Horava-Lifshitz gravitly  \cite{hor}  and its variant \cite{naka} we propose the following action
\ben \nonumber S&=&\frac{2}{\kappa^2}\int dtd^{3}xN\sqrt{h}\left(K_{ij}K^{ij}- \lambda K^2\right)+\int dtd^{3}xN\sqrt{h}\left[\frac{\kappa^2\mu^2(\Lambda_{W}R-3\Lambda^{2}_{W})}{8(1-3\lambda)}\right]+\nonumber\\
&&+\int dtd^{3}xN\sqrt{h}\left(\alpha NB^{ij}K_{ij}\right)+\int dtd^{3}xN\sqrt{h}L_{m},\een
where $K_{ij}$ defines the extrinsic curvature of surfaces of constant time in the spacetime foliation
\be K_{ij}=\frac{1}{2N}(\dot{h}_{ij}-\nabla_{i}N_{j}-\nabla_{j}N_{i})\ee
with the lapse field $N$ and the shift variable $N_i$ defined in the metric in the ADM decomposition
\be ds^2=-N^2dt^2+h_{ij}\left(dx^{i}+N^{i}dt\right)\left(dx^{j}+N^{j}dt\right),\ee
where $i=1,2,3$. One should also define the following effective quantities: the effective Newton constant, the effective speed of light and effective cosmological constant, respectively 
\be G=\frac{\kappa^2}{32\pi c}, \qquad c=\frac{\kappa^2\mu}{4}\sqrt{\frac{\Lambda_{W}}{1-3\lambda}}, \qquad  \Lambda=\frac{3}{2}\Lambda_{W}.\ee
Let us now explicit the anisotropic scaling of the principal  ingredients of the theory with dynamical critical exponent $z$ in $d$ spatial dimensions as follows
\be\label{scales-E} [dt]={-z}, \quad
 [dx]={-1},\quad [c]={z-1}, \quad
 [\mu]=1, \quad
 [\Lambda_{W}]=2, \quad
 [K_{ij}]=[\dot{h}_{ij}]={z},\quad
  [\kappa]=\frac{z-d}{2},\quad
[E]=z.\ee
In the proposed action we keep only linear terms in $B^{ij}$ which has the interpretation of being proportional to a gradient flow equation in systems with detailed balance \cite{hor}. The dimensional analysis reveals that it scales with the dimension
\be [B^{ij}]=d\ee
and since
\be [H]=z, \qquad [\rho]={z+d}\ee
it is natural to consider
\be
 [B^{ij}]=\left[\frac{\rho}{H}\right]=d.
\ee
For $d=3$ and $z=3$ the dimension of $B^{ij}$ coincides with dimension of energy (\ref{scales-E}).

If this tensor has dimension of energy we attempt to recognize it  as the energy of a {\it non-conventional fluid} different from the conventional matter and energy which fills all the 
three-dimensional spatial region with volume $V\propto H^{-d/z}$ of our Universe and couples to gravity through  $B^{ij}$. 

Let us now postulate the tensor $B^{\mu\nu}$ as follows
\be B^{\mu\nu}=T^{\mu\nu}V=\frac{\Omega_{3}}{H}T^{\mu\nu},\ee
whose spatial part reads
\be B^{ij}=\frac{\Omega_{3}}{H}T^{ij}=\frac{4\pi}{3H}T^{ij},\ee
where $T^{\mu}_{\nu}={\rm diag}(\rho, p, p, p)$ is the energy-momentum tensor of this fluid.

Adopting the spatially flat, homogenous and isotropic `cosmological metric'
\be ds^2=-N^2(t)dt^2+a^2(t)dx_{i}dx^{i},\qquad i=1,2,3, \ee
where $^{(3)}R=0$, we find
\be B^{ij}=\frac{4\pi}{3H}\frac{1}{a^2}p\,\delta{ij},\qquad
K^{ij}=\frac{1}{N}\frac{H}{a^2}\delta^{ij}, \qquad H=\frac{\dot{a}(t)}{a(t)}\qquad i=1,2,3, \ee
and
\be K=\frac{3}{N}H, \qquad
 K_{ij}K^{ij}=\frac{3}{N^2}H^2.\ee
Now the proposed Lagrangian can be written as
\ben\label{sch} S=-\frac{1}{\kappa^2}\int dtd^{3}xN\sqrt{h}\left[\frac{6(3\lambda-1)}{N^2}H^2\right]+\int dtd^{3}xN\sqrt{h}\left[\frac{3\kappa^2\mu^2\Lambda^{2}_{W}}{8(3\lambda-1)}\right]\nonumber\\
+\int dtd^{3}xN\sqrt{h}(8\pi\alpha p)+\int dtd^{3}xN\sqrt{h}L_{m}.\een
The third term of Eq.~(\ref{sch}) is similar to that  proposed in Schutz formalism \cite{sch}.  Now we take the following variation 
\be\frac{\delta S}{\delta N}=\sqrt{h}\frac{6(3\lambda-1)}{\kappa^2N^2}H^2+\sqrt{h}\left(\frac{4c^2}{\kappa^2}\Lambda\right)+\sqrt{h}(8\pi\alpha p)-N\sqrt{h}\rho_{m}=0,\ee
where we define the conventional matter and energy density as
\be \rho_{m}=-\frac{1}{N\sqrt{h}}\frac{\delta S_{m}}{\delta N}.\ee
Assuming the variation at  $N=1$ we find the important equation
\be\label{master-eq}
\frac{6(3\lambda-1)}{\kappa^2}H^2=-\frac{4c^2}{\kappa^2}\Lambda-8\pi\alpha p+\rho_{m}.\ee
Let us make a first consideration. If the aforementioned fluid is itself the vacuum energy we just arrive at
\be p=-\frac{\Lambda c}{8\pi G}=-\frac{4c^2}{\kappa^2}\Lambda,\ee
Now and in the next discussions we shall take  $\alpha=1/8\pi$. Thus, we find
\ben\frac{6(3\lambda-1)}{\kappa^2}H^2&=&\frac{4c^2}{\kappa^2}(\Lambda-\Lambda)+\rho_{m}\nonumber\\
&=&\rho_{m}.\een
We conclude that in this scenario the cosmological constant exerts no gravitational effect regardless of its value \cite{carr}.

\section{Construction out of Horava-Melby-Thompson gravity }
\label{HMT-sec}

Although in our previous analysis we focus entirely on the first Horava gravity version, it is now well-known that there is a generalization of this theory due to Horava and Melby-Thompson \cite{Horava:2010zj} --- See also \cite{daSilva:2010bm} for related issues. In this section we show that there is a close relationship between our previous study and Horava-Melby-Thompson (HMT) gravity. In  the HMT gravity the authors consider a theory with larger symmetry in order to solve the scalar graviton problem. The action is basically given by 
\ben \nonumber S&=&\frac{2}{\kappa^2}\int dtd^{3}xN\sqrt{h}\left(K_{ij}K^{ij}- \lambda K^2\right)+\frac{2}{\kappa^2}\int dtd^{3}xN\sqrt{h}{ V}+\nonumber\\
&&+\frac{2}{\kappa^2}\int dtd^{3}xN\sqrt{h}\,\nu \theta^{ij}\left(K_{ij}+\frac12\nabla_i\nabla_j\nu\right)-\int dtd^{3}x\sqrt{h}\,A(R-2\Omega)+\int dtd^{3}xN\sqrt{h}\tilde{\cal L},\een
where $\nu$ is the Newtonian prepotential with dimension $[\nu]=z-2$ and $A$ is a gauge field that plays the role of the Newton potential with dimension $[A] = 2z - 2$.  The potential $V$ is unconstrained by the new symmetries just as in the minimal theory \cite{Horava:2010zj}.  The new coupling constant $\Omega$ with dimension $[\Omega]=2$ has the same dimension of $\theta^{ij}$ that is defined as 
\ben
 \theta^{ij}=R^{ij}-\frac12g^{ij}R+\Omega g^{ij}
\een
and $\tilde{\cal L}=\tilde{\cal L}(N,N_i,\nu,\chi,g_{ij},A)$ is a general Lagrangian that include matter fields $\chi$. Let us now choose the $\nu$-dependence as follows
\ben
\tilde{\cal L}(\nu)=\frac{4\pi F(t)}{N}\ln{\left(\frac{\nu}{\nu_0}\right)},
\een
such that 
\ben
\frac{\delta S}{\delta \nu}=0\qquad\longrightarrow\qquad F(t)=p.
\een
On the other hand, if we simply assume $V=A=0$, $\nu\equiv\nu(t)$ and identify  $\nu\theta^{ij}=(-{4\pi}/{3H})T^{ij}$, that is equivalent to our previously $B^{ij}$, we have
\be\frac{\delta S}{\delta N}=\sqrt{h}\frac{6(3\lambda-1)}{\kappa^2N^2}H^2-\sqrt{h}p-N\sqrt{h}\rho=0,\ee
where we define the {\it total} energy density as
\be \rho=-\frac{1}{N\sqrt{h}}\frac{\delta \tilde{S}}{\delta N}.\ee
Assuming the variation at  $N=1$ we find the important equation
\be\label{master2-eq}
\frac{6(3\lambda-1)}{\kappa^2}H^2= p+\rho.\ee
Since for all species we should have 
\ben
\rho=\rho_\Lambda+\rho_{\rm matter}+\rho_{\rm radiation}+... \nonumber\\ p=p_\Lambda+p_{\rm matter}+p_{\rm radiation}+...
\een
being $p_\Lambda=-\rho_\Lambda$ the {\it cosmological constant equation of state}, then from equation (\ref{master2-eq}) its gravitational contribution is completely cancelled out regardless of its value. Several other cosmological issues on this theory was addressed  in \cite{Wang:2010wi}.

\section{Cosmological constant plus non-conventional fluid}
\label{s2}

Let us now consider a cosmological scenario with the non-conventional fluid in terms of only two-species ($\rho_m=0$) with a portion composed of vacuum energy and another part called $p_f$
\be p=-\frac{4c^2}{\kappa^2}\Lambda+p_{f},\ee
that we insert into Eq.~(\ref{master2-eq}) to obtain
\be\label{pf-part}\frac{6(3\lambda-1)}{\kappa^2}H^2=p_{f}.\ee
Notice that, as expected, once again theres is no gravitational effect coming from the cosmological constant. 
The equation for the acceleration of the part $p_f$ of the non-conventional cosmological fluid can be found by substituting its equation of state $p_f=\omega\rho_f$ into Eq.~(\ref{pf-part}) and differentiating it with respect to time to obtain
\be\frac{12(3\lambda-1)}{\kappa^2}H\dot{H}=\omega\dot{\rho}_{f}\ee
Now with the help of the  energy conservation equation 
\be \dot{\rho}_{f}+3H(\rho_{f}+p_{f})=0,\ee
we find
\be\frac{4(3\lambda-1)}{\kappa^2}\dot{H}=-\omega (1+\omega)\rho_{f}.\ee
By developing derivatives in $H$ and substituting Eq.~(\ref{pf-part}) we just find 
\be\frac{4(3\lambda-1)}{\kappa^2}\left[\frac{\ddot{a}}{a}-\frac{\kappa^2}{6(3\lambda-1)}\omega\rho_{f}\right]=-\omega(1+\omega)\rho_{f},\ee
or simply by putting in the `familiar' form reads
\be\label{acele}\frac{\ddot{a}}{a}=-\frac{\kappa^2}{12(3\lambda-1)}\omega(1+3\omega)\rho_{f}.\ee

According to the modified Friedmann equation (\ref{pf-part}), for  $\lambda>1/3$  we should have $\omega>0$. In this case the Eq.~(\ref{acele}) leads to decelerating Universe ($\ddot{a}<0$). On the other hand, for $\lambda<1/3$ we must have $\omega<0$ and the Eq.~(\ref{acele}) implies that  for $\ddot{a}>0$ we must have $\omega(1+3\omega)>0$ that means
\be -1< \omega<-\frac{1}{3}.\ee
This corresponds to an accelerating phase of the Universe even if the cosmological constant does not play a role in the evolution.

\section{Cosmological constant plus conventional matter as a fraction of the non-conventional fluid}
\label{s3}

As in the previous section, let us now consider a cosmological scenario with the non-conventional fluid in terms of only two-species with a portion composed of vacuum energy and another part as a fraction describing the conventional matter in terms of the pair ($\eta p_m$, $\epsilon\rho_m)$. Thus, substituting 
\be p=-\frac{4c^2}{\kappa^2}\Lambda+ \eta p_{m},\ee
into Eq.~(\ref{master2-eq}) we obtain
\be\label{fried2}\frac{6(3\lambda-1)}{\kappa^2}H^2= \eta p_{m}+\epsilon\rho_{m},\ee
and using the equation of state $p_m=\omega\rho_m$ for a conventional matter (or energy) we get to the following modified Friedmann equation
\be\label{fried-2nd}\frac{6(3\lambda-1)}{\kappa^2}H^2=(\eta\omega+\epsilon)\rho_{m}.\ee
Notice that, as expected, in this case there is no gravitational effect coming from the cosmological constant either. 

As in the previous case, the equation for the acceleration of the conventional fraction of the non-conventional cosmological fluid can be found by differentiating the modified Friedmann equation (\ref{fried-2nd}) with respect to time to obtain 
\be\frac{12(3\lambda-1)}{\kappa^2}H\dot{H}=(\eta\omega+\epsilon)\dot{\rho}_{m}.\ee
Now using the energy conservation equation for the conventional matter
\be \dot{\rho}_{m}+3H(\rho_{m}+p_{m})=0,\ee
we find
\be\frac{4(3\lambda-1)}{\kappa^2}\dot{H}=-(\eta\omega+\epsilon)(1+\omega)\rho_{m}.\ee
By developing derivatives in $H$ and substituting Eq.~(\ref{fried-2nd})  we just find 
\be\frac{4(3\lambda-1)}{\kappa^2}\left[\frac{\ddot{a}}{a}-\frac{\kappa^2}{6(3\lambda-1)}(\eta\omega+\epsilon)\rho_{m}\right]=-(\eta\omega+\epsilon)(1+\omega)\rho_{m},\ee
or in the more familiar form reads
\be\frac{\ddot{a}}{a}=-\frac{\kappa^2}{12(3\lambda-1)}f(\omega)\rho_{m},\ee
where $f(\omega)=3\eta\omega^2+(3\epsilon+\eta)\omega+\epsilon $. In the regime that $\eta\to1$ and $\epsilon\to0$, we simply recover the pure non-conventional cosmological fluid previously studied. 

Let us now study the possible cosmological scenarios  according to the sign of $f(\omega)$. It is easy to check that 
\ben
&&f(\omega)>0, \qquad \mbox{for $\omega<-{\epsilon}/{\eta}$ or $\omega>-\frac{1}{3}$},\nonumber\\
&&f(\omega)<0, \qquad \mbox{for $-{\epsilon}/{\eta}<\omega<-\frac{1}{3}$}.
\een

We shall now discuss the following scenarios. Firstly, for  $\lambda>1/3$ the Friedmann equation (\ref{fried-2nd}) implies that  
$\omega>-\epsilon/\eta$, which for $\epsilon,\eta>0$ allows a region with $\omega$ negative.  In this case the sign of $f(\omega)$ determines only an accelerating phase ($f(\omega)<0$) of the Universe. For decelerating phase ($f(\omega)>0$) of Universe one has to approach the ratio of the parameters $\epsilon/\eta$ to $1/3$. 

Secondly, for  $\lambda<1/3$ we know from the Friedmann equation (\ref{fried-2nd}) that now $\omega<-\epsilon/\eta$,  which for $\epsilon,\eta>0$ allows $\omega$ always negative. Now the sign of $f(\omega)$ determines an accelerating phase ($f(\omega)>0$) or decelerating phase ($f(\omega)<0$) of Universe provided that the ratio $\epsilon/\eta=1/3$. 

\section{conclusions}
\label{conc}

In this paper we have construct a self-tuning model out of the Horava-Lifshitz (HL) gravity. We have shown that its possible to find a perfect cancelation of the cosmological constant due to a tensor field $B^{ij}$. As a consequence the theory gives a modified Friedmann equation whose conventional or non-conventional fluid contribution comes also from the pressure and not only from the density as in the conventional general relativity. This makes all the difference in constructing the self-tuning model. Despite of the cosmological constant cancelation, there are still conventional cosmological scenarios present in the model, composing the accelerating and decelerating phases of the Universe. Since our model with deformed HL gravity can be seen as a sector of the HMT gravity it is expected that there are no issues concerning the scalar graviton present in the original HL gravity.

\acknowledgements

We would like to thank CNPq, CAPES, PNPD/PROCAD - CAPES for partial financial support.


\end{document}